\documentclass[reprint,showpacs,preprintnumbers,amsmath,amssymb,prc,floatfix]{revtex4-1}
\usepackage{float}
\usepackage{color}
\usepackage{graphicx}% Include figure files
\usepackage{dcolumn}% Align table columns on decimal point
\usepackage{bm}% bold math
\usepackage{xcolor}
\usepackage{multirow}
\usepackage{comment}
\usepackage[colorlinks,citecolor=blue,linkcolor=red,anchorcolor=blue,filecolor=blue,urlcolor=blue]{hyperref}
\begin{document}
\title{Cluster decay dynamics of Actinides yielding non-Pb-daughter}

\author{Joshua T. Majekodunmi$^1$}
\email{majekjoe1@gmail.com}
\author{M. Bhuyan$^{2}$}
\email{bunuphy@um.edu.my}
%\author{K. Anwar$^1$}
%\author{N. Abdullah$^1$}
\author{Raj Kumar$^3$}
\email{rajkumar@thapar.edu}

%%%%%%%%%%%%%%%%%
\affiliation{$^1$Institute of Engineering Mathematics,  Universiti Malaysia Perlis, Arau, 02600, Perlis, Malaysia}
%%%
\affiliation{$^2$Center for Theoretical and Computational Physics, Department of Physics, Faculty of Science, University of Malaya, Kuala Lumpur 50603, Malaysia}
\affiliation{$^3$School of Physics and Materials Science, Thapar Institute of Engineering and Technology, Patiala, Punjab 147004, India}
%%
%\affiliation{$^4$Institute of Research and Development, Duy Tan University, Da Nang 550000, Vietnam}
%%%%%%%%%

\date{\today}% It is always \today, today,

\begin{abstract}
\noindent
The cluster dynamics of radioactive nuclei decaying to neighbouring daughter nuclei of the double magic $^{132}$Sn and $^{208}$Pb is investigated using the relativistic mean-field (RMF) approach with NL3$^*$ parameter set within the preformed cluster-decay model (PCM). The novel feature of the present study is the application of the newly derived preformation formula, laying the groundwork for accessing the break-up of the Q-value: preformation energy,  cluster emission energy and the recoil energy of the daughters formed. The energy associated with cluster preformation is theoretically quantified for the first time. This treatment underscores the shell effect, pairing correlation as well as the blocking of particular orbitals by unpaired nucleons. To ascertain the applicability of the new formula, the PCM based calculations are carried out with nuclear potential obtained using the phenomenological M3Y and microscopic RMF-based R3Y nucleon-nucleon (NN) potentials along with corresponding densities. We found a marginal variation that can be attributed to the difference in their barrier properties, however, the predictions for the case of both M3Y and R3Y potentials are found to agree well with the experimental half-lives. Although none of the considered reaction systems yields a double magic daughter nucleus, we found that the kinematics of their cluster emissions is governed by their proximity to the shell closure. The deduced systematic of the recoil energy in cluster decays can provide valuable insight for the synthesis of elements in superheavy mass region in the future. \\
\par 
\end{abstract}
%%%%%%%%%%%%%%%%%%%
\maketitle

%%%%%%%%%%%%%%%%%%%%%%%%%%%%%%%%%%%%%
\section{INTRODUCTION} \label{intro}
\noindent
The bedrock of nuclear physics can be traced to the discovery of natural radioactivity. Particularly, the spontaneous disintegration of nuclei in which the emitted particles are heavier than $^4$He but lighter than fission fragments was first predicted by Sandulescu \cite{sand80} and experimentally observed by Rose and Jones \cite{rose84}. Thereafter, the possibility of emission of several other clusters ranging from $^{14}$C $-^{34}$Si from various heavy nuclei have been discovered \cite{bone07} and their daughter nuclei are usually the double magic nucleus $^{208}$Pb and its neighbouring nuclei. This dominant double magic structure in the heavier fragment strongly indicates the shell effect on cluster radioactivity. Thus, cluster emission for heavy nuclei is a highly asymmetry fission process \cite{deni13}.  
%%%

Beside the empirical relations \cite{jain22,path21} used for the calculation of decay half-lives, the literature is replete with several microscopic descriptions of cluster emission \cite{lova98,ward11,ward18,math19}.  Clustering occurs as an intermediate process between $\alpha$-decay and spontaneous fission. From the theoretical front, it follows the description of the Gamow model of $\alpha$-decay which hinges on the quantum tunnelling effect \cite{maru74,fink74}. Nonetheless, cluster decay models can be classified into two different approaches based on their treatment of cluster emissions. The first is the fission-like models \cite{poen85,poen86,poen02,poen10} where it is assumed that clusters are gradually formed as the parent nucleus undergoes successive geometrical deformation until it reaches a saddle or scission point. The second refers to the alpha-like models e.g. the preformed cluster-decay model (PCM)  \cite{gupt88, maje22,yahy22,josh22} which holds the assumption that clusters are pre-formed within the decaying parent nucleus before its penetration across the interaction barrier. Unlike the fission-like models where the preformation probability $P_0$ is usually taken as unity (i.e. $P_0=1$), the PCM requires the calculation of a realistic $P_0$ which could be challenging due to the complexities associated with the many-body system and the variability of the nuclear potential. As such, certain range of $P_0$ values have been assigned to different regions of the nuclear chart (in Table 2 of Ref.\cite{ahme15} and the references therein)  and various $P_0$ formulae \cite{blend88,sant21}  have been employed to reproduce the experimental half-lives. Yet, no complete/clear link that is traceable to the mechanism of cluster emission has been established.
%%%

However, we have demonstrated in our recent studies \cite{maje22L,maje23NPA,maje23EPhys} that a thorough estimate of cluster preformation can shed new light on the essential features of the decaying system. Further, the cluster preformation energy was systematically quantified for the first time. We realised that there is a distinct but subtle difference in the behaviours of the radioactive nuclei yielding the double magic daughter $^{208}$Pb in comparison and those with daughters in its vicinity. This could be due to numerous reasons (e.g pairing and odd-even staggering effects associated with open-shell nuclei, relatively higher binding energies (and Q-values) of parent nuclei with double magic daughters at shell closure or a relatively low number of valence particles of isotopes near the shell closure and other factors)  which the present study aims to establish in connection with their effect(s) on the preformation and mechanism of cluster emissions. Moreover, the effective number of valence particles (or holes) is known to largely influence the parameterization of several nuclear quantities \cite{cast85}. Besides, in $\alpha$-decay studies \cite{ahme15,ni09}, it has been demonstrated that the preformation probability of nuclei near magic numbers is susceptible to rapid variations. As an extension, the present study will be devoted to probing the behaviour and trend of $P_0$ for near magic-number daughters in cluster decays. Thus, cluster formation energy and the recoil effect of the daughter nuclei are taken into consideration within the PCM. The only adjustable parameter of the PCM is the neck-length $\Delta$R which accounts for the relative separation (whose proximity potential limit is in the range ($0.0-2.0$) fm  \cite{bloc77}) between the decay fragments and decides the $1^{st}$ turning point of the barrier penetration \cite{kuma12}. 

It is assumed that the preformed clusters tunnels across the barrier jointly built by the Coulomb and the nuclear potentials. The nuclear potential between the emitted cluster and the daughter nuclei is obtained from the well-established double-folding technique \cite{satc79}. One of the necessary inputs for the double-folding technique is the nuclear matter densities of the decay fragment which is estimated in this study from the relativistic mean-field (RMF) \cite{bogu77,rein89,gamb90,sero86} with the NL$3^*$ parameter set \cite{lala09}. The RMF formalism is well known for its successful description of various bulk properties of nuclei in both ground and excited states \cite{Bisw20,itag20,pani21}. The second input for the double-folding process is the effective nucleon-nucleon (NN) interaction \cite{satc79}. Here, we employ the phenomenological M3Y NN potential and the microscopic R3Y NN potential \cite{maje22,josh22,maje22i} which stems from the non-linear RMF Lagrangian \cite{bisw15,sahu14} for the sake of comparison.  In our previous studies \cite{kuma12,maje22,maje22i}, we have demonstrated that $\Delta$R = 0.5 fm  is suitable enough for M3Y interaction while the R3Y fits nicely at $\Delta$R = 1.0 fm in cluster radioactivity. The Q-values are estimated from the experimental binding energies given in a recent mass table \cite{wang21}. The penetration probability $P$ is calculated from the well-known Wentzel-Kramers-Brillouin (WKB) approximation. 

Sec. \Ref{theory} briefly describes the theoretical framework: non-linear RMF Lagrangian, the double-folding technique for both R3Y and M3Y potentials, the preformed cluster-decay model (PCM) and our recently derived $P_0$ formula and its components. The obtained results are discussed in Sec. \Ref{result}. Finally, the conclusion and summary of this work are given in Sec. \Ref{summary}.
%%%%%%%%%%%%%%%%%%%%%%%%%

\section{Theoretical formalism}
\label{theory} 
\noindent 

The relativistic mean-field (RMF) approach denotes a kind of implementation of the density functional theory based on a Lorentz covariance. An atomic nucleus is considered as a system composed of Dirac nucleons, exchange various mesons ($\sigma$, $\omega$ and $\rho$) and the photon field ($A_\mu$) through an effective Lagrangian given by \cite{josh22,maje22,bogu77,rein89,gamb90,sero86},
%%%%
\begin{eqnarray}
{\cal L}&=&\overline\psi_i\left\{i\gamma^{\mu}\partial_{\mu}-M\right\}\psi_i+\frac{1}{2}\partial^{\mu}\sigma\partial_{\mu}\sigma\nonumber\\
&&-\frac{1}{2}m_{\sigma}^{2}\sigma^2-\frac{1}{3}g_2\sigma^3-\frac{1}{4}g_3\sigma^4-g_\sigma\overline{\psi}_i\psi_i\sigma\nonumber\\
&&-\frac{1}{4}\Omega^{\mu\nu}\Omega_{\mu\nu} +\frac{1}{2}m^2_\omega \omega^\mu \omega_\mu-g_\omega\overline{\psi}_i\gamma^\mu\psi_i \omega_\mu\nonumber\\
&&-\frac{1}{4}\vec B^{\mu\nu}.\vec B_{\mu\nu}+\frac{1}{2}m^2_\rho\vec \rho^\mu.\vec \rho_\mu-g_\rho\overline{\psi}_i\gamma^\mu\vec{\tau}\psi_i.\vec \rho^\mu\nonumber\\
&&-\frac{1}{4}F^{\mu\nu}F_{\mu\nu}-e\overline{\psi}_i\gamma^\mu(\frac{1-\tau_{3i}}{2})\psi_i A_\mu. 
\label{lag}
\end{eqnarray}
%%%%%
The parameters $g_\sigma$, $g_\omega$, $g_\rho$ denotes the respective coupling constants of the mesons whose corresponding masses are $m_\sigma$, $m_\omega$ and  $m_\rho$ while  $M$ is the mass of nucleons. Similarly,  $g_2$, $g_3$ and $\frac{e^2}{4\pi}$ are the coupling constants of the non-linear terms. The third component of the isospin is $\tau_{3i}$. Here, the contribution of the $\pi$-meson has been omitted in Eq. (\Ref{lag}) in the mean-field calculation due to its pseudoscalar nature \cite{ring96,sero86}. A detailed description of the field tensors for $\omega^{\mu}$, $\vec \rho_\mu$ and $A_\mu$ fields can be found in Ref \cite{sing22} and the references therein. By taking the field tensors  as classical fields, the Dirac equation is obtained for the nucleons and simplified as,
\begin{equation}
[-i\alpha.\nabla+\beta(M^*+g_\sigma\sigma)+g_\omega\omega+g_\rho\tau_3\rho_3]\psi_i=\epsilon_i\psi_i. 
\end{equation}
Similarly, the Klein-Gordon equations for the participating mesons are simplified as
\begin{eqnarray}
 (-\nabla^2+m^2_\sigma)\sigma(r)&=&-g_\sigma\rho_s(r)-g_2\sigma^2(r)-g_3\sigma^3(r),\nonumber\\
   (-\nabla^2+m^2_\omega)V(r)&=&g_\omega\rho(r),\nonumber\\
   (-\nabla^2+m^2_\rho)\rho(r)&=&g_\rho\rho_3(r).
\end{eqnarray}
These equations are solved self consistently using the NL3$^*$ parameter set. Within the limit of one-meson exchange for a heavy and static baryonic medium, the microscopic R3Y NN potential is obtained as, 
\begin{eqnarray}
V_{eff}^{R3Y}(r)&=&\frac{g^2_\omega}{4\pi}\frac{e^{-m_\omega r}}{r}+\frac{g^2_\rho}{4\pi}\frac{e^{-m_\rho r}}{r}-\frac{g^2_\sigma}{4\pi}\frac{e^{-m_\sigma r}}{r}\nonumber\\
&&+\frac{g^2_2}{4\pi}re^{-2m_\sigma r} +\frac{g^2_3}{4\pi}\frac{e^{-3m_\sigma r}}{r} +J_{00}(E)\delta(s).
\label{r3y} 
\end{eqnarray}
%%%%%%%%
Here $J_{00}(E)\delta(s)$ is the zero-range pseudopotential symbolizing the exchange effect. Eq. (\Ref{r3y}) is similar to the phenomenological prescription of Reid-Elliott \cite{satc79} called   M3Y NN potential which is constructed to reproduce the G-matrix element. The  M3Y NN potential takes the form,
\begin{equation}
V_{eff}^{M3Y}(r)=7999\frac{e^{-4r}}{4r}-2134\frac{e^{-2.5r}}{2.5r}+J_{00}(E)\delta(s).
\label{m3y}
\end{equation}
%%%%%
The double folding technique \cite{satc79} is employed to calculate the nuclear interaction potential $V_n(R)$ and expressed as
%%%%
\begin{equation}
    V_n(R)=\int dr_c\int dr_d \rho_c(\vec r_c)\rho_d(\vec r_d)V_{eff}(\vec r_{cd}=\vec R+\vec r_d- \vec r_c),
    \label{fold}
\end{equation}
%%%%
where $\rho_c$ and $\rho_d$ are the nuclear densities of the cluster and daughter nuclei respectively. $V_n(R)$ given by Eq. (\Ref{fold}) combines with the Coulomb potential $V_C(R)=\frac{Z_{c}Z_d}{R}e^2$ to obtain the total interaction potential
\begin{equation}
    V(R)= V_n(R)+V_C(R)+V_\ell(R),
\end{equation}
which is used to estimate the WKB penetration probability  and hence, the cluster decay half-lives using the preformed cluster-decay model (PCM) \cite{kuma12}.  Note that the contribution of the centrifugal potential $V_\ell(R)=\frac{\hbar^2\ell(\ell+1)}{2\mu R^2}$ (where $\mu=m(A_cA_d/A)$ is the reduced mass) is neglected in the ground state to ground state transitions where the angular momentum $\ell=0$. The penetration probability of clusters across the tunnelling path is  given as
\begin{equation}
    P=P_aW_iP_b, \label{pen}
\end{equation}
which involves a three step process \cite{maje22}. Here,
\begin{eqnarray}
    P_a&=&\exp\left(-\frac{2}{\hbar}\int^{R_i}_{R_a}\{2\mu[V(R)-V(R_i)]\}^{1/2}dR\right),\\
    \nonumber \mbox{and}\\
    P_b&=&\exp\left(-\frac{2}{\hbar}\int_{R_i}^{R_b}\{2\mu[V(R_i)-Q]\}^{1/2}dR\right). 
    \label{wkb} 
\end{eqnarray}
and $W_i$ in Eq. (\Ref{pen}) is estimated as unity, following the Greiner and Scheid de-excitation ansatz \cite{grei86}. 

%%%%%%%%%%%%%
\begin{ruledtabular}
\begin{table}
\caption{\label{tab1} Fitting parameters a, b and c for the preformation formula in Eq. (\ref{P0}) for known experimentally favoured cluster decays. The Chi-square ($\chi^2$) for the half-life predictions of the M3Y and R3Y interactions are given in columns 5 and 6 respectively. Note that odd-odd cluster emitters have not been experimentally observed.%\\ \textcolor{red}{Follow the modified footnote of the table and check-in the text accordingly.}
}
    \centering
    \begin{tabular}{|c |c|c|c|c|c|}
         System &\multicolumn{3}{c|}{Constant Parameters}& \multicolumn{2}{c|}{ $\chi^2$}
         \\
         \cline{2-6}
         & a&b&c&M3Y&R3Y\\
         \hline  			
         {\it e-e} &17.67&0.114&8.00&0.284&0.075\\
         {\it o-A} &16.12&0.119&0.88$^\dag$&0.390$^\P$&0.063$^\P$\\
          &16.12&0.119&4.02$^\S$&0.061 &0.065 \\  
    \end{tabular}
    \label{tab 1}
\footnotesize{$^\dag$Parameter $c$ appears lower for systems having daughters with non-magic neutron numbers ($N_d \leq 126$). }\\
\footnotesize{$^\S$Systems having daughters with a magic neutron (or neighbours)  require a higher value of parameter $c$ $(126\leq N_d\leq 128)$. }\\
\footnotesize{$^\P$Experimental lower limits are available here.}  
\end{table}
\end{ruledtabular}

%%%%%%%%%%%%%%
%%%%%
\subsection{Preformed cluster-decay model (PCM)}
Within the  preformed cluster-decay model (PCM), the half-life $T_{1/2}$ (and decay constant $\lambda$) is usually expressed in terms of  the penetration probability $P$, and preformation probability $P_0$ as, 
%%%%
\begin{equation}
    T_{1/2} =\frac{\ln2}{\lambda}, \hspace{0.5cm} \lambda= \nu_{0} P_0 P. \label{11}
\end{equation}
The assault frequency $\nu_0$ has nearly constant value of $10^{21}$ s$^{-1}$ and can be calculated as 
\begin{equation}
    \nu_{0}=\frac{\mbox{ velocity }}{R_0}=\frac{\sqrt{2E_{c}/\mu}}{R_0}, \label{afreq}
\end{equation}
%%%%%
where $R_0$ symbolises the radius of the parent nucleus and $E_{c}$ is the kinetic energy of the emitted cluster.  The Q-values are calculated from the experimental binding energies data \cite{wang21}  using the expression
 \begin{equation}
     Q=(BE_d+BE_c)-BE_p, \label{qval}
 \end{equation}
where $BE_p$, $BE_d$ and $BE_c$ are the binding energies of the parent, daughter nuclei and the emitted cluster respectively.
%%%

As a result, we have studied the relationship among various theoretically established properties/factors that influences cluster preformation such as the cluster mass $A_c$ \cite{sing11}, mass and charge asymmetries $\eta_A=(A_d-A_c)/(A_d+A_c)$ and $\eta_Z=(Z_d-Z_c)/(Z_d+Z_c)$ \cite{sing11}, the relative separation between the centers of the  fragments $r_B=1.2(A_c^{1/3}+A_d^{1/3})$ \cite{deli09,qian12} and the Q-value \cite{isma14}. Thus, the newly proposed $P_0$ formula \cite{maje22L} is of the form
\begin{equation}
     \log P_0 = -\frac{aA_c\eta_A}{r_B}-Z_c\eta_Z+bQ+c. \label{P0}
\end{equation}\\
Here $a$, $b$ and $c$ are the fitting parameters in Table \Ref{tab 1}. A thorough and intuitive analysis of the third term on the right-hand side of Eq. (\Ref{P0}) reveals the fractional amount of the decay energy contributed during the cluster pre-formation process. Thus a narrow bridge linking the contributions of the decay energy is constructed via the new $P_0$ formula. Thus, the Q-value is presented in terms of its disbursement in the kinematics of cluster emission as,
%%%
\begin{eqnarray}
Q=\overbrace{\underbrace{ bQ}_{\substack{\text{energy  }\\ \text{contributed in}\\ \text{cluster formation}}}+\underbrace{\kappa\sqrt{Q}}_{\substack{\text{energy }\\ \text{contributed in}\\ \text{cluster emission}}}}^{E_c}+\underbrace{E_d}_{\substack{\text {recoil}\\ \text{energy of}\\ \text{daughter nucleus}}} \label{qexpd}
\end{eqnarray}
%%%%
where the $\kappa\sqrt{Q}$ is the \textit{energy contributed in cluster emission}. Further, the kinetic energy of the emitted cluster of Gupta \textit{et al.} \cite{sing11} is simplified as
\begin{eqnarray}
 E_c=\frac{A_d}{A}Q=bQ+\kappa\sqrt{Q},
 \label{ke}
\end{eqnarray}
which yields
\begin{equation}
    \kappa=\sqrt{Q}\left(\frac{A_d}{A}-b\right). \label{kap}
\end{equation}
The quantity $\kappa$ in Eq.(\ref{kap}) refers to the tunneling factor. The newly derived expressions in Eq.s (\Ref{P0})-(\Ref{kap}) are analysed and  explained in section \Ref{result}.

%%%%%%%%%%%%
\begin{figure*}
\includegraphics[scale=0.32]{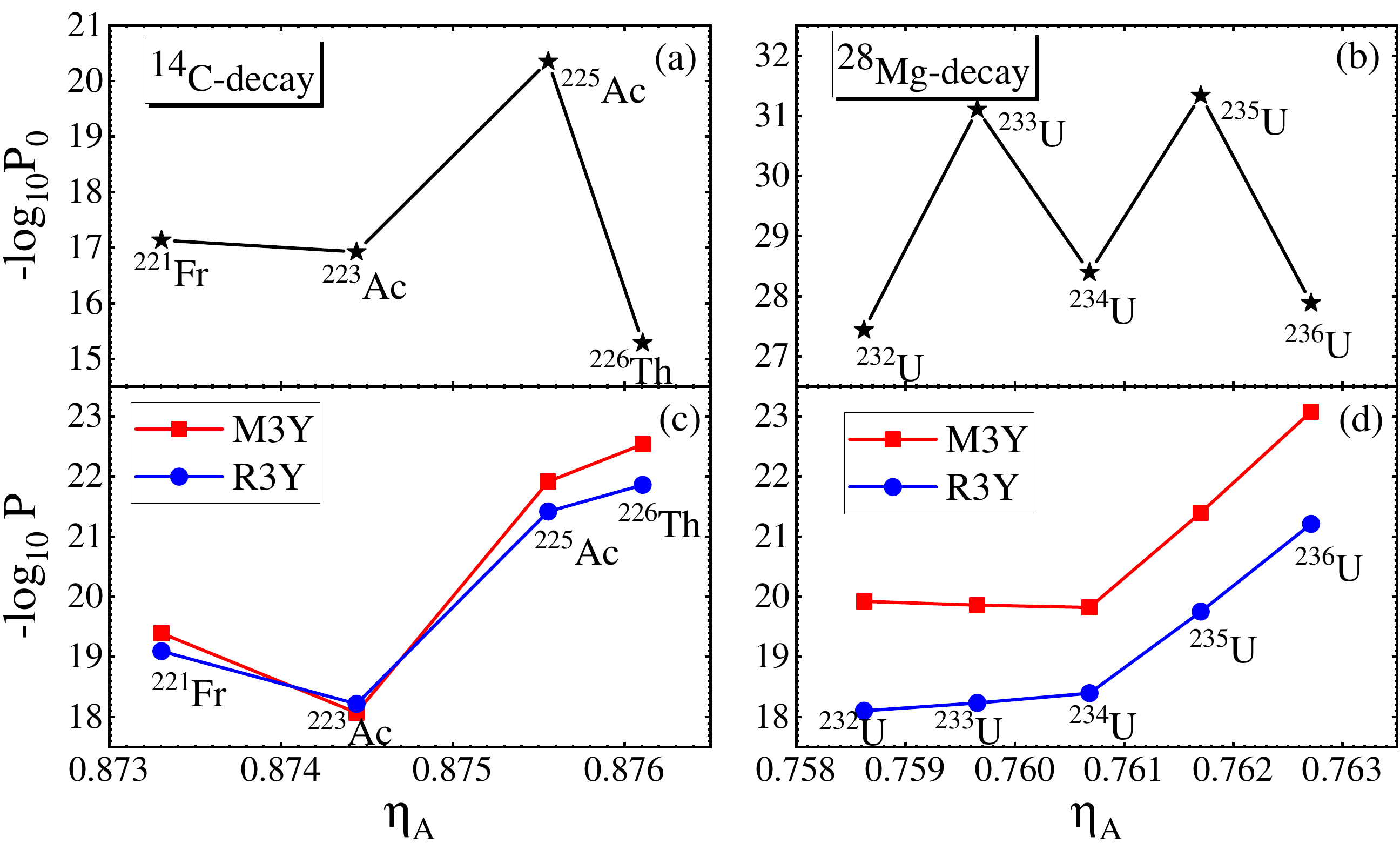}% 
\caption{\label{fig 1} The calculated preformation probability $P_0$ (in logarithmic scale) versus the mass asymmetry $\eta_A$ in Eq. (\ref{m_asy}) for  (a) $^{14}$C emission from different actinides and  panels (b) $^{28}$Mg emission from $^{232-236}$U isotopes. The profiles of the penetration probability $P$ (in logarithmic scale)  for both cases are given in panels (c) and (d) respectively. }
\end{figure*}
%%%%%%%%%%%%%%
\begin{table*}
\caption{\label{tab2} The penetrability, decay half-lives and mass asymmetry of various actinides using the M3Y and R3Y NN potentials. The Q-values are calculated from the experimental binding energies given in a recent mass table \cite{wang21}. The experimentally observed half-lives are taken from Ref. \cite{bone07}.}
\begin{ruledtabular}
\begin{tabular}{cccccccccc}
Parents 	&Emitted	&Daughters	&Q-value 	&\multicolumn{2}{c}{Penetrability}		&\multicolumn{3}{c}{$\log_{10}T_{1/2}$}	&$\eta_A$		\\
\cline{5-6} \cline{7-9}
nuclei&cluster&nuclei&(MeV)&P$^{M3Y}$	&P$^{R3Y}$&T$^{exp}_{1/2}$	&T$^{M3Y}_{1/2}$	&T$^{R3Y}_{1/2}$& \\
\hline
$^{114}$Ba	&$^{12}$C	&$^{102}$Sn	&19.02	&1.08$\times 10^{-19}$	&3.08$\times 10^{-18}$	&$>4.10$	&11.90	&10.44	&0.789\\
$^{221}$Fr	&$^{14}$C	&$^{207}$Tl	&31.29	&4.11$\times 10^{-20}$	&8.04$\times 10^{-20}$	&14.52	&14.90	&14.61	&0.873\\
$^{223}$Ac	&$^{14}$C	&$^{209}$Bi	&33.06	&5.28$\times 10^{-19}$	&6.05$\times 10^{-19}$	&12.6	&13.57	&13.51	&0.874\\
$^{225}$Ac	&$^{14}$C	&$^{211}$Bi	&30.48	&1.82$\times 10^{-22}$	&3.84$\times 10^{-22}$	&17.16	&17.34	&20.16	&0.876\\
$^{226}$Th	&$^{14}$C	&$^{212}$Po	&30.55	&5.21$\times 10^{-23}$	&1.39$\times 10^{-22}$	&$>15.3$	&15.95	&15.53	&0.876\\
$^{230}$Th	&$^{24}$Ne	&$^{206}$Hg	&57.76	&2.85$\times 10^{-22}$	&2.98$\times 10^{-21}$	&24.61	&24.91	&23.89	&0.791\\
$^{231}$Pa	&$^{24}$Ne	&$^{207}$Tl	&60.41	&2.08$\times 10^{-20}$	&8.48$\times 10^{-20}$	&23.23	&23.62	&23.01	&0.792\\
$^{232}$U	&$^{28}$Mg	&$^{204}$Hg	&74.32	&1.61$\times 10^{-20}$	&7.89$\times 10^{-19}$	&$>22.26$	&25.58	&23.89	&0.759\\
$^{233}$U	&$^{28}$Mg	&$^{205}$Hg	&74.23	&1.62$\times 10^{-20}$	&5.83$\times 10^{-19}$	&$>27.59$	&29.24	&27.69	&0.760\\
$^{234}$U	&$^{28}$Mg	&$^{206}$Hg	&74.11	&1.53$\times 10^{-20}$	&4.01$\times 10^{-19}$	&25.14	&25.64	&24.23	&0.761\\
$^{235}$U	&$^{28}$Mg	&$^{207}$Hg	&72.43	&5.13$\times 10^{-22}$	&1.77$\times 10^{-20}$	&$>28.09$	&30.98	&29.44	&0.762\\
$^{235}$U	&$^{29}$Mg	&$^{206}$Hg	&72.48	&4.79$\times 10^{-22}$	&2.55$\times 10^{-21}$	&$>28.09$	&28.55	&27.82	&0.753\\
$^{236}$U	&$^{28}$Mg	&$^{208}$Hg	&70.73	&1.38$\times 10^{-23}$	&6.18$\times 10^{-22}$	&27.58	&29.11	&27.46	&0.763\\
$^{236}$U	&$^{30}$Mg	&$^{206}$Hg	&72.27	&2.11$\times 10^{-22}$	&3.70$\times 10^{-21}$	&27.58	&29.22	&27.97	&0.746\\
$^{238}$U	&$^{34}$Si	&$^{204}$Pt	&85.01	&8.95$\times 10^{-22}$	&1.29$\times 10^{-20}$	&29.04	&30.63	&29.47	&0.714\\
$^{237}$Np	&$^{30}$Mg	&$^{207}$Tl	&74.79	&5.70$\times 10^{-21}$	&4.72$\times 10^{-20}$	&$>26.93$	&27.91	&26.99	&0.747\\
$^{238}$Pu	&$^{32}$Si	&$^{206}$Hg	&91.19	&1.19$\times 10^{-18}$	&6.51$\times 10^{-18}$	&25.27	&25.62	&24.88	&0.731\\
$^{240}$Pu	&$^{34}$Si	&$^{206}$Hg	&91.06	&8.74$\times 10^{-19}$	&1.83$\times 10^{-18}$	&$>25.52$	&27.07	&26.75	&0.717\\
$^{241}$Am	&$^{34}$Si	&$^{207}$Tl	&93.96	&1.46$\times 10^{-17}$	&1.17$\times 10^{-17}$	&$>22.71$	&25.61	&25.71	&0.718\\
$^{252}$Cf	&$^{46}$Ar	&$^{206}$Hg	&126.75	&6.02$\times 10^{-15}$	&2.79$\times 10^{-16}$	&$>15.89$	&26.81	&28.14	&0.635\\
$^{252}$Cf	&$^{48}$Ca	&$^{204}$Pt	&137.97	&3.68$\times 10^{-15}$	&3.68$\times 10^{-14}$	&$>15.89$	&26.81	&25.81	&0.619\\
$^{252}$Cf	&$^{50}$Ca	&$^{202}$Pt	&138.32	&7.14$\times 10^{-15}$	&6.92$\times 10^{-16}$	&$>15.89$	&27.03	&28.05	&0.603 
\end{tabular}
\end{ruledtabular}
\end{table*}
%%%%%%

\section{CALCULATIONS AND DISCUSSIONS}
\label{result}
In the context of heavy fragment nuclear decay (otherwise called cluster radioactivity), the parent nucleus having mass $A$ undergoes a spontaneous disintegration in which two separate fragments (daughter nucleus and the emitted cluster with masses $A_d$ and $A_c$ respectively) are formed. The mass asymmetry between these fragments is expressed as \cite{tava07,gupt10}
\begin{equation}
    \eta_A=\frac{A_d-A_c}{A_d+A_c}=1-\frac{2A_c}{A}. \label{m_asy}
\end{equation}
It is worth noting that for cluster decays, $0.60\leq\eta_A\leq0.90$  whereas $0.92\leq\eta_A\leq0.97$ in alpha decay and $\eta_A\leq0.5$ in fission studies \cite{tava07}. Interestingly, all $\eta_A$ values of the considered reaction systems are found within the established range for cluster decays as shown in the last Column of Table \Ref{tab2}.

%%%%%%%%%%%%%%%%%%%%%
\begin{table*}
\caption{\label{tab3} The predicted driving potential $d_{pot}=V(R_a)-Q$ %barrier properties (barrier height $V_B$ and barrier lowering $\Delta V_B=V_B-Q$) 
from M3Y and R3Y interactions are given in columns (2-5). The preformation properties calculated from a newly derived set of Eqs. (\Ref{P0}) - (\Ref{kap}) are given in columns (6-11). }
\begin{ruledtabular}
\begin{tabular}{ccccccccccc}
Reaction 	&\multicolumn{4}{c}{Barrier Properties}	&\multicolumn{6}{c}{Preformation Properties}\\
\cline{2-5} \cline{6-11} \\
systems&$V^{M3Y}(R_a)$	&$d_{pot}^{M3Y}$	&$V^{R3Y}(R_a)$	&$d_{pot}^{R3Y}$&$P_0$	&b.Q	&$\kappa$	&$\kappa\sqrt{Q}$	&$E_c$	&$E_d$\\
 & (MeV)&(MeV)&(MeV)&(MeV)& Eq. (\Ref{P0})&(MeV)& & (MeV)&(MeV)&(MeV)\\
\hline \\
$^{114}$Ba$\rightarrow ^{12}$C	+ $^{102}$Sn	&32.37	&13.34	&29.56	&10.54	&2.61$\times 10^{-15}$	&2.17	&3.40	&14.85	&17.02	&2.00\\
$^{221}$Fr$\rightarrow ^{14}$C	+ $^{207}$Tl	&42.49	&11.20	&36.73	&5.44	&7.32$\times 10^{-18}$	&3.74	&4.57	&25.57	&29.31	&1.98\\
$^{223}$Ac$\rightarrow ^{14}$C	+ $^{209}$Bi	&43.63	&10.57	&37.82	&4.76	&1.19$\times 10^{-17}$	&3.95	&4.70	&27.04	&30.99	&2.08\\
$^{225}$Ac$\rightarrow ^{14}$C	+ $^{211}$Bi	&42.38	&11.91	&36.38	&5.90	&4.40$\times 10^{-21}$	&3.64	&4.52	&24.94	&28.58	&1.90\\
$^{226}$Th$\rightarrow ^{14}$C	+ $^{212}$Po	&42.87	&12.32	&36.82	&6.27	&5.17$\times 10^{-16}$	&3.49	&4.55	&25.16	&28.66	&1.89\\
$^{230}$Th$\rightarrow ^{24}$Ne	+ $^{206}$Hg	&75.30	&17.54	&66.43	&8.67	&1.00$\times 10^{-25}$	&6.60	&5.94	&45.13	&51.73	&6.03\\
$^{231}$Pa$\rightarrow ^{24}$Ne	+ $^{207}$Tl	&76.40	&15.99	&67.52	&7.11	&2.61$\times 10^{-26}$	&7.22	&6.04	&46.92	&54.13	&6.28\\
$^{232}$U$\rightarrow ^{28}$Mg	+ $^{204}$Hg	&95.30	&20.98	&85.26	&10.94	&3.64$\times 10^{-28}$	&8.49	&6.60	&56.86	&65.35	&8.97\\
$^{233}$U$\rightarrow ^{28}$Mg	+ $^{205}$Hg	&94.74	&20.51	&84.64	&10.42	&7.85$\times 10^{-32}$	&8.87	&6.55	&56.44	&65.31	&8.92\\
$^{234}$U$\rightarrow ^{28}$Mg	+ $^{206}$Hg	&94.18	&20.07	&84.02	&9.90	&4.02$\times 10^{-29}$	&8.47	&6.59	&56.77	&65.24	&8.87\\
$^{235}$U$\rightarrow ^{28}$Mg	+ $^{207}$Hg	&93.35	&20.93	&83.04	&10.62	&4.59$\times 10^{-32}$	&8.65	&6.48	&55.15	&63.80	&8.63\\
$^{235}$U$\rightarrow ^{29}$Mg	+ $^{206}$Hg	&91.33	&18.86	&80.32	&7.84	&1.36$\times 10^{-29}$	&8.66	&6.45	&54.88	&63.53	&8.94\\
$^{236}$U$\rightarrow ^{28}$Mg	+ $^{208}$Hg	&92.52	&21.79	&82.06	&11.33	&1.30$\times 10^{-28}$	&8.08	&6.45	&54.26	&62.34	&8.39\\
$^{236}$U$\rightarrow ^{30}$Mg	+ $^{206}$Hg	&92.44	&20.17	&81.56	&9.29	&6.72$\times 10^{-30}$	&8.26	&6.45	&54.83	&63.08	&9.19\\
$^{238}$U$\rightarrow ^{34}$Si	+ $^{204}$Pt	&106.25	&21.24	&93.90	&8.89	&6.03$\times 10^{-32}$	&9.71	&6.85	&63.15	&72.87	&12.14\\
$^{237}$Np$\rightarrow ^{30}$Mg	+ $^{207}$Tl	&93.76	&18.97	&82.84	&8.05	&5.02$\times 10^{-30}$	&8.93	&6.52	&56.39	&65.32	&9.47\\
$^{238}$Pu$\rightarrow ^{32}$Si	+ $^{206}$Hg	&110.02	&18.83	&98.48	&7.30	&4.40$\times 10^{-30}$	&10.42	&7.17	&68.51	&78.93	&12.26\\
$^{240}$Pu$\rightarrow ^{34}$Si	+ $^{206}$Hg	&109.32	&18.26	&96.91	&5.84	&2.15$\times 10^{-31}$	&10.41	&7.10	&67.76	&78.16	&12.90\\
$^{241}$Am$\rightarrow ^{34}$Si	+ $^{207}$Tl	&110.85	&16.89	&98.39	&4.43	&3.68$\times 10^{-31}$	&11.22	&7.17	&69.48	&80.70	&13.26\\
$^{252}$Cf$\rightarrow ^{46}$Ar	+ $^{206}$Hg	&142.31	&15.56	&127.19	&0.44	&5.77$\times 10^{-35}$	&14.48	&7.92	&89.13	&103.61	&23.14\\
$^{252}$Cf$\rightarrow ^{48}$Ca	+ $^{204}$Pt	&158.79	&20.82	&143.59	&5.62	&9.13$\times 10^{-35}$	&15.77	&8.17	&95.92	&111.69	&26.28\\
$^{252}$Cf$\rightarrow ^{50}$Ca	+ $^{202}$Pt	&156.11	&17.79	&139.07	&0.76	&2.89$\times 10^{-35}$	&15.81	&8.08	&95.07	&110.87	&27.44\\
\end{tabular}
\end{ruledtabular}
\end{table*}
%%%%%%
%%%%%%%%%%%%
\begin{figure}
\includegraphics[scale=0.6]{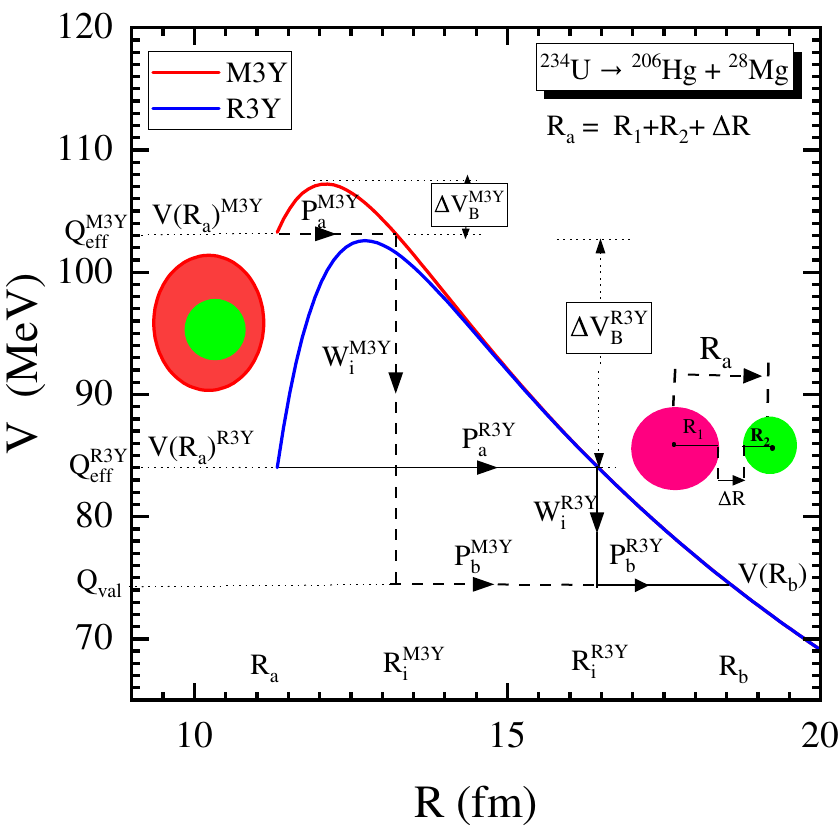}% 
\caption{\label{sp} The total interaction potential of $^{234}$U $\rightarrow ^{206}$Hg + $^{28}$Mg as a function of the mass-center distance between the decaying fragments (R). Prior to the decay process, the emitted cluster $^{28}$Mg (green circle) is assumed to pre-exist within the $^{234}$U -parent nuclei (red circle). After the quantum tunneling process, a neck-length $\Delta$R is formed between  the decay fragments ($^{28}$Mg and $^{205}$Hg-daughter nucleus (Pink colour)). The black dash-lines and black solid lines are used to depict the three-step barrier penetration process for M3Y and R3Y respectively.}
\end{figure}
%%%%%%%%%%%%%%

%%%%%%%%%%%%%%%%%%%%%%%%%%%
\begin{figure*}
\includegraphics[scale=0.28]{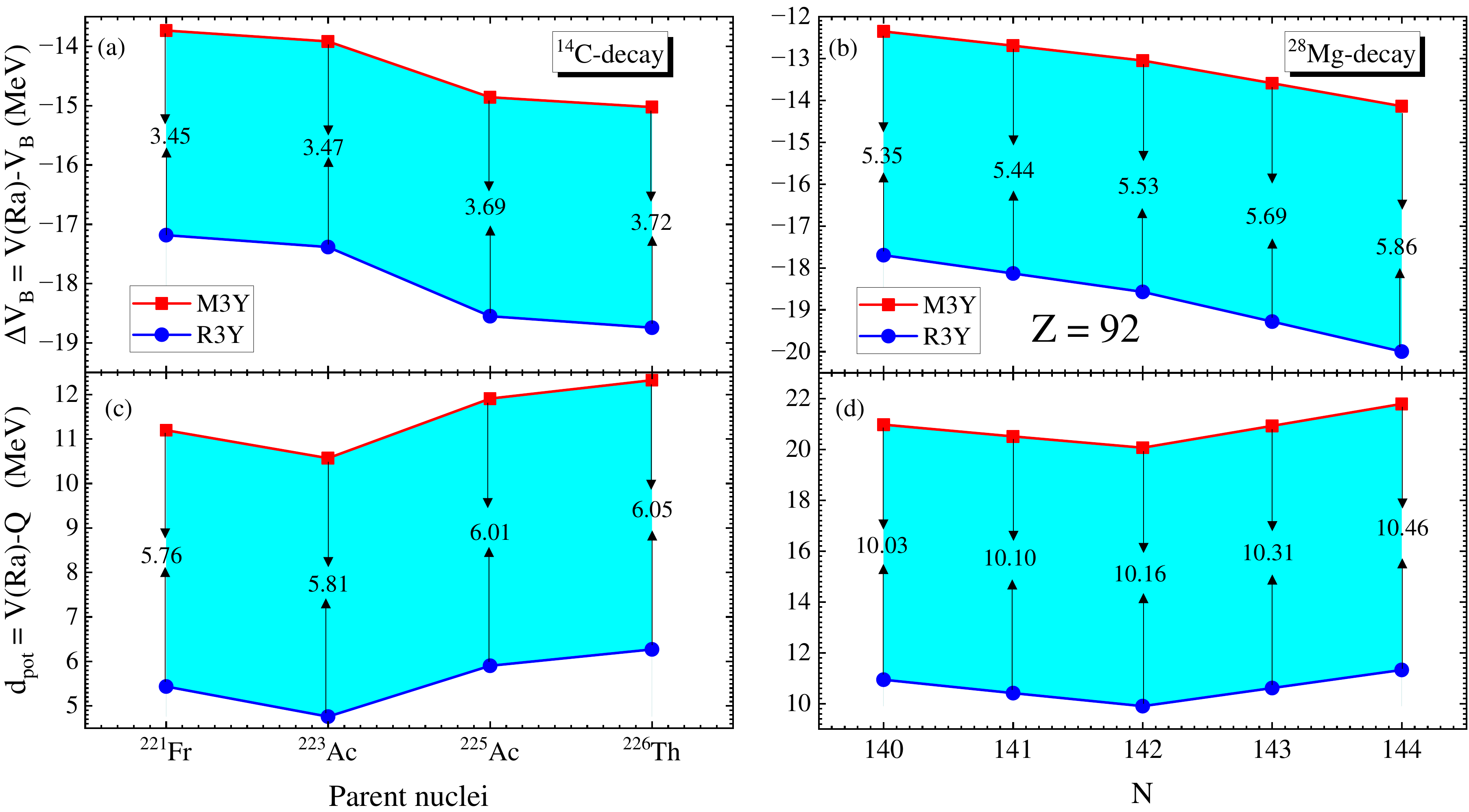}% Here is to import EPS art
\caption{\label{fig 2} Variation of the barrier lowering $\Delta V_B$  for (a) $^{14}C$ decay from various actinides and (b)$^{A}U\rightarrow ^{28}Mg+^{A-28}Hg$. The profile of the driving potential $d_{pot}=V(R_{a})-Q$  for both cases are shown in 
 (c) and (d) respectively.}
\end{figure*}
%%%%%%%
%%%%%%%%%
\begin{figure}
\includegraphics[scale=1.05]{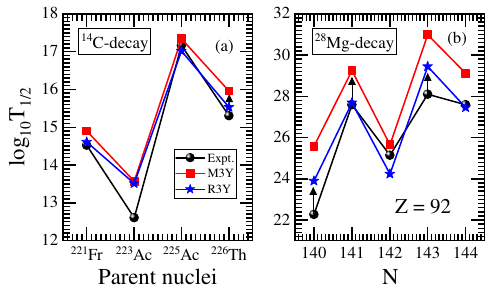}% Here is to import EPS art
\caption{\label{fig HL} Logarithmic half-lives of (a) different actinides emitting $^{14}$C cluster and (b) Uranium isotopes emitting $^{28}$Mg clusters ($^{A}U\rightarrow ^{28}Mg+^{A-28}Hg$).}
\end{figure}
%%%%%%%%%%%%
%%%%%%%%%
\begin{figure}
\includegraphics[scale=0.4]{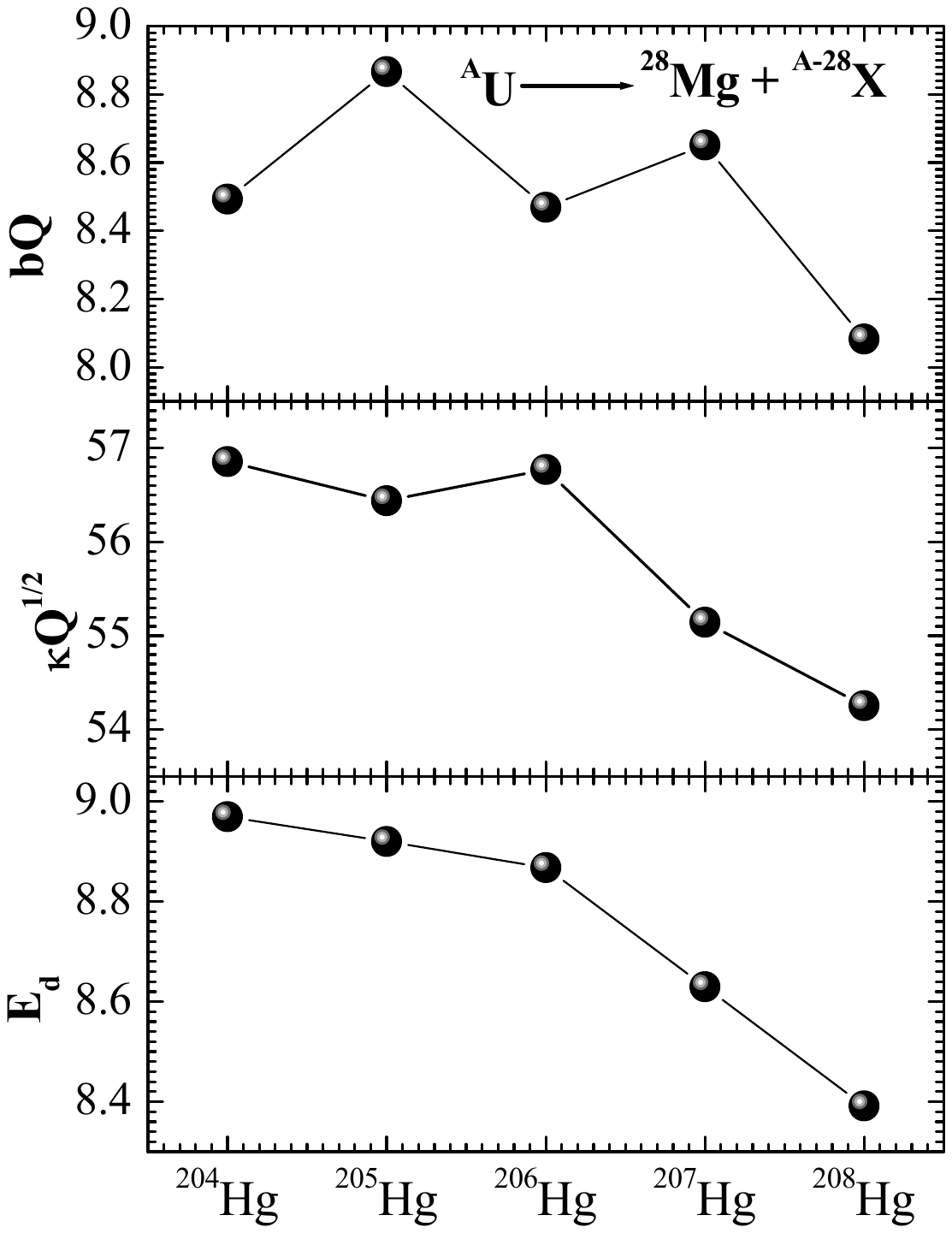}% Here is to import EPS art
\caption{\label{fig 3} Variation of the preformation properties:  (a) weighted Q-values of various  Uranium isotopes as a function of the neutron number of the daughters formed from the emission of $^{28}$Mg clusters respectively, (b) cluster emission energy and (c) recoil energy of the daughter nuclei.}
\end{figure}
%%%%%%%%%%%%

Fig. \ref{fig 1} displays the variation of the preformation probability and penetration probability (in logarithmic scale) as a function of the mass asymmetry of the parent nuclei $(\eta_A)$. It is worth noting that the preformation probability $(P_0)$ and penetration probability $(P)$ are moderately model-dependent quantities.  The calculated $P_0$ values from Eq. (\Ref{P0}) and the actual values are given in Column 6 of Table \Ref{tab3}. The upper panel of the figure shows the preformation of (a) $^{14}$C cluster emission from different parent nuclei. From the figure, a slight drop is noticed in the magnitude of $P_0$ between $^{221}$Fr (yielding $^{207}$Tl $N_d=126, Z_d=81$ daughter) and $^{223}$Ac that yields $^{209}$Bi of $N_d=126, Z_d=83$. Thus, the small dip can be correlated with the neutron magic shell closure. However, the effect of an enhanced clusterization is visible as $P_0$ increases drastically for an odd-A isotope of larger $\eta_A $ i.e $^{225}$Ac resulting in the formation of daughters (with $N_d=128$) having two loosely bonded neutrons above the shell, taking part in the cluster formation process. Nonetheless, this effect may not be so apparent in medium-mass nuclei \cite{deng16}. The usual trend is maintained with increasing mass asymmetry. A closer look at the figure suggests that pairing plays a vital role in cluster preformation and even-even nuclei may face less structural hindrance as compared to those from odd-parents. Details of the correlation of ground state spin (in terms of paired neutrons and orbital filling) and its effect on preformation can be found in Ref. \cite{isma13}.
%%%%
 
Fig \Ref{fig 1} (b) illustrates the effect of pairing correlation between the {\it even-even} and {\it odd-A} parent  Uranium isotopic chain with each following a similar trend. As the considered nuclei are laying at and/or near the $\beta-$stable region of the nuclear chart, hence the BCS approach \cite{gamb90} is reasonable to predict the appropriate pairing effect, and it can be expected that the prediction will be not changed by adopting more appropriate approach such as Bogoliubov transformation \cite{ring96}. The lower panel of the figure shows the profile of the penetration probability for (c) different nuclei and (d) $^{232}$U isotopes. In both cases, it is obvious that the penetration probability of systems is not merely influenced by their masses but also, by the collective nature of the daughters formed. Hence, the dip formed at $^{223}$Ac and $^{234}$U is illustrative of the proximity to a magic proton number of daughter $N_d$ produced as explained earlier. The small variation in the  M3Y and R3Y predictions can be attributed to the difference in their individual barrier properties. 
%%%%%

The anatomy of the three-step barrier penetration process is further typified in Fig. \ref{sp}. As a representative sample, the figure shows the total interaction potential of  $^{234}$U $\rightarrow ^{206}$Hg + $^{28}$Mg as a function of the internuclear distance between the decaying fragments. Prior to the decay process, the emitted cluster $^{28}$Mg (green circle) is assumed to pre-exist within the $^{234}$U -parent nuclei (red circle). After the quantum tunneling process, a neck-length $\Delta$R is formed between  the decay fragments (the emitted cluster $^{28}$Mg and $^{205}$Hg-daughter nucleus (Pink colour). The black dash-lines and black solid lines are used to depict the three-step barrier penetration process for M3Y and R3Y respectively. The process begins at point $R_a$ whose corresponding potential is $V(R_a)$ (also known as the effective Q-value, $Q_{eff}$), usually higher than the $Q$-value. Thereafter, de-excitation begins at point $R_i$ which is primarily decided by the peculiarity and profile of the employed NN potential. For R3Y, $R_i$ is further extended due  to its repulsive nature as compared to its cognate M3Y NN potential. In each case, the de-excitation probability $W_i=1$, following the Greiner and Scheid's ansatz \cite{grei86} and then penetration process continues until it reaches point $R_b$ having a potential $V(R_b)=Q$.  

The difference between $V(R_a)$ and the barrier height $V_B$ basically provides the barrier lowering parameter $\Delta V_B$, which gives a vivid picture of the behaviour of the microscopic R3Y and phenomenological M3Y interaction. 
Fig. \Ref{fig 2} (a) and (b) illustrates the variation of the barrier lowering  $\Delta V_B$  for $^{14}C$ decay from various actinides and $^{A}U\rightarrow ^{28}Mg+^{A-28}Hg$ respectively, with respect to the mass (and neutron) number of their respective parent nuclei. The shaded portion in the figures depicts the difference between the $\Delta V_B^{M3Y}$ (red solid square) and $\Delta V_B^{R3Y}$ (blue solid circle). A close assessment of both figures reveals that the M3Y and R3Y potentials manifest similar trend, although at different magnitudes. This is due to the repulsive nature of the R3Y interaction potential, as earlier mentioned. Particularly, in Fig. \Ref{fig 2}(a), $\Delta V_B$ is found to decrease significantly with each increase in the mass of the considered actinides emitting the same ($^{14}$C) cluster. A similar phenomenon is noticed in Fig \Ref{fig 2}(b). However,  across the  uranium 
 ($Z=92$) isotopic chain, the decline in $\Delta V_B$ is found to arise primarily from the increase in the neutron number (N). The M3Y and R3Y barrier properties within the PCM can be examined more closely by probing the driving potential $d_{pot}=V(R_{a})-Q$. Usually, the $d_{pot}$ is  influenced by the choice of interaction potential as well as the nature of the decaying parent nucleus and decay fragments (cluster and daughter nucleus). Fig. \Ref{fig 2}(c)-(d) depicts the profile of $d_{pot}$  as a function of the mass (and neutron number) of the parent nuclei for $^{14}C$ decay from various actinides and $^{A}U\rightarrow ^{28}Mg+^{A-28}Hg$ respectively. In Fig. \Ref{fig 2}(c), both M3Y and R3Y reproduced the deepest minima at  $^{223}$Ac  corresponding to $^{209}$Bi ($Z=83, N=126$) daughter nucleus i.e just above the proton shell closure and at the neutron shell closure. Thereafter, the driving potential rises with each increase in the mass number of the parent nuclei. Also, a similar occurrence is found in Fig. \ref{fig 2}(d) where the minima is formed at $N=142$ corresponding to $^{206}$Hg ($Z=80, N=126$) i.e at neutron shell closure along the isotopic chain.
 
As one examines the shaded portion  of Fig. \Ref{fig 2}(a)-\Ref{fig 2}(d), it is salient to note that the difference between the M3Y and R3Y predictions reduces successively for parent nuclei with lower masses. Thus, we presume that the predictions from the  M3Y and R3Y may quantitatively agree for parent nuclei with relatively lower mass such that $\Delta V_B^{M3Y}-\Delta V_B^{R3Y}\approx 0$. In other words, the relative difference in the M3Y and R3Y interactions could be mass region-dependent. We have recently shown \cite{maje23EPhys} that a wider difference may ensue for studies involving heavy particle radioactivity (HPR).

%%%%%
 Fig. \Ref{fig HL} (a) and (b) depicts the logarithmic half-lives of different actinides emitting $^{14}$C and $^{A}U\rightarrow ^{28}Mg+^{A-28}Hg$. In both cases, the $\log_{10}T_{1/2}$ predictions of M3Y (red solid square) and R3Y (blue star) are in good agreement with the experimentally measured half-lives (black sphere) and the lower limit (black sphere with an upward arrow) for $^{226}$Th. As expected, the lowest minima is formed at the nearest double magic neighbour $^{223}$Ac having $^{209}$Bi ($Z=83, N=126$) daughter nucleus in Fig. \Ref{fig HL}(a). It is worth noting that the precise experimental half-lives of most of the studied systems in Fig \Ref{fig HL}(b) are unavailable but the predictions here, especially for $N=140$ are most probable since, by principle, the lowest minima is expected at $142$, whose daughter is formed at the neutron magic number $N=126$. The half-lives of all the systems considered in this work are given in Columns 7 - 10 of Table \Ref{tab2}. In most cases, the predictions of the microscopic R3Y interaction, give a relatively closer agreement with the experiment \cite{bone07}. Beside the effect of the neutron shell closures of the daughter nuclei, the $\log_{10}T_{1/2}$ values are generally lower for parent nuclei with even mass numbers,  which is reminiscent of an odd-even staggering behaviour.
%%%%

The third term of Eq. (\Ref{P0}) is the fractional amount of the decay energy contributed during the cluster formation process only. The contribution of the Q-value to each stage of the kinematics of cluster emission is fully spelt out in Eq. (\Ref{qexpd}) and thus, their qualitative estimate is given in Fig. \Ref{fig 3} for $^{A}U\rightarrow^{28}Mg +^{A-28}Hg$. Particularly,  Fig. \Ref{fig 3} typifies the share of energy participating in the cluster preformation of each reaction system as a function of the neutron number of the daughter nuclei. The weighted Q-value i.e. $bQ$ can only be influenced by the decay energy and parameter $b$. For the sake of accuracy, the Q-values are calculated from the experimental binding energy data \cite{wang21}. A detailed inspection of the figure shows that the magnitude of $bQ$ is all-time higher for odd-A nuclei as compared to their neighbouring even-even nuclei. This presupposes that the energy required for the preformation of nuclear clusters can be relatively higher for {\it odd-A} than those with {\it even-even} parents. This behaviour reflects the odd-even staggering effects originating from both pairing correlations as well as the blocking of particular orbitals by unpaired nucleons associated with $^{205}$Hg and $^{207}$Hg. Interestingly, this trend is repeated for all the systems under study in Table \Ref{tab3} and our observation corroborates with recent findings \cite{seif15,yang22}.
%%%%

Although none of the considered reaction systems yields a double magic daughter nucleus, it is apparent that the kinematics of their cluster emission is governed by their proximity to the shell closure. A vivid picture of this fact is given in the footnote of Table \Ref{tab1} where the parameter $'c'$ assumes a uniform fitting to either $N_d\leq126$ or $126\leq N_d \leq 128$ for all odd-A nuclei. The effect of parameter $'c'$ is further revealed in Fig. \Ref{fig 3}(b) where $\kappa\sqrt{Q}$ (the energy required for the emission of a preformed cluster) is plotted as a function of the neutron number of the participating daughter nuclei. In the figure, a conspicuous minimal is formed at $N_d=125$  $(N_d\leq126$) corresponding to a relatively low parameter $'c'$. In the same vein,  at $N=127$, there is a nearly imperceptible bend arising from the larger value of $'c'$ since it is found within the range $126\leq N_d \leq 128$. Thus, the mass of the parent nucleus as well as those of the decay fragment plays a decisive role in the tunnelling factor $\kappa$. By definition,  $\kappa$ is the fraction of the $Q$-value required, just for the propagation of the preformed cluster. Its actual value is given in column 8 of Table \Ref{tab3} where it becomes evident that there is a direct proportionality between the quantity $\kappa$ and $bQ$ i.e $bQ$ increases with increasing value of $\kappa$. This implies that there is a close correlation between the amount of energy contributed during cluster preformation and its tunnelling as portrayed in Eq. (\Ref{qexpd}).
%%%%
 
Fig. \Ref{fig 3} (c) shows the variation of the recoil energy as a function of the neutron number of the daughter $N_d$. The recoil energy maintains a regular profile until it attains the neutron magic shell closure $N_d=126$ where a conspicuous shift is noticed in the isotopic chain. Again, this indicates the dominance of the shell effect. Detailed discussions on the shell closure effect as well as its correlation with the isotopic shift in charge radius and the single-particle energy levels can be found in Ref. \cite{bhu21}. A notable inference that can be drawn from the figure is that the recoil energy of a reaction system decreases as the mass of the corresponding daughter (and parent) nucleus increases along the isotopic chain (e.g. $^{A}$U $\rightarrow^{28}$Mg $+^{A-28}$Hg) in which the same cluster is emitted provided that no constituent (proton and neutron number) of the daughter formed is a magic number.On the other hand, a careful inspection of the last column of Table \Ref{tab3} shows that the recoil energy increases proportionately for the reaction systems with the same daughter nuclei but of different clusters and parent nuclei (with increasing size/mass). For example \\
$^{221}$Fr$\rightarrow ^{14}$C	+ $^{207}$Tl\hspace{3mm} yields $E_d=1.98$ MeV,\\
$^{231}$Pa$\rightarrow ^{24}$Ne	+ $^{207}$Tl\hspace{3mm} yields $E_d=6.28$ MeV,\\
$^{237}$Np$\rightarrow ^{30}$Mg	+ $^{207}$Tl\hspace{3mm} yields $E_d=9.47$ MeV,\\
$^{241}$Am$\rightarrow ^{34}$Si	+ $^{207}$Tl\hspace{3mm} yields $E_d=13.26$ MeV.\\
In other words, $E_d$ increases proportionately with systems yielding the same daughter nucleus but of increasing cluster masses if and only if no constituent of the daughter formed is a magic number. This observation is also true for the emission of clusters of different masses from the same parent nucleus. For example:\\
$^{238}$Pu$\rightarrow ^{32}$Si	+ $^{206}$Hg \hspace{3mm} yields $E_d=12.26$ MeV, \\
$^{240}$Pu$\rightarrow ^{34}$Si	+ $^{206}$Hg \hspace{3mm} yields $E_d=12.90$ MeV, \\
provided that the daughter formed are non-double magic nuclei. We hope that the analysis of the  recoil energy in cluster emissions will be informative for meaningful extrapolations in the (synthesis of the) superheavy region since recent studies have shown that cluster decay could be a dominant decay mode in the superheavy region \cite{ward18,math19,poen12,sant22}.
%%%%%%%%%%%

\section{SUMMARY AND CONCLUSIONS}
\label{summary}
The dynamics of cluster emission are studied within the relativistic mean-field (RMF) formalism using the NL3$^*$ parameter set. Assuming that a cluster pre-exists as an entity within the parent nucleus, we have applied our newly developed cluster preformation $P_0$ formula for radioactive nuclei decaying to yield daughters in the vicinity of double magic shell closure. The $P_0$ formula opens a novel route for remodelling the $Q-$value such that it gives a quantitative estimate of the energy contributed during the cluster preformation process, transmission energy of the preformed clusters and the recoil energy of the daughters formed. Besides, to ensure the applicability of the formula, we have employed the well-known M3Y and microscopic-based R3Y NN potentials for the analysis. Despite the difference in the barrier properties of these NN potentials, our result reveals that with the inclusion of the new $P_0$ formula, the calculated half-lives are in good agreement with the experimentally measured half-lives. We have also demonstrated that the kinematics of cluster emission is governed by the proximity of the corresponding daughter nuclei to the shell closure. The pairing correlation and the odd-even staggering effect arising from the unpaired neutrons are appraised. A detailed analysis of the recoil energy that can be extrapolated for the forthcoming synthesis of superheavy nuclei is also discussed. However, in principle, the shapes degree of freedom of each of the participating nuclei plays a crucial role in its description. This is an interesting future problem that will be given due consideration.
%%%%

\section*{acknowledgments}
The authors would like to acknowledge the support from the Fundamental Research Grant Scheme (FRGS) under the grant number FRGS/1/2019/STG02/UNIMAP/02/2 from the Ministry of Education Malaysia stipulated with the Institute of Engineering Mathematics (IMK),  UniMAP as the beholder. This work was supported by FOSTECT Project Code: FOSTECT.2019B.04, FAPESP Project Nos. 2017/05660-0, and Science Engineering Research Board (SERB), File No. CRG/2021/001229. \\

\end{document}